\documentclass{PoS}
\usepackage{epsf}

\newcommand{\half}{\mbox{\small $\frac{1}{2}$}}          
\newcommand{\msbar}{\mbox{\tiny $\overline{MS}$}}        
\newcommand{\ripmom}{\mbox{\tiny $R\!I^\prime\!\!-\!\!M\!O\!M$}} 
\newcommand{\rimom}{\mbox{\tiny $R\!I\!\!-\!\!M\!O\!M$}}
\newcommand{\rgi}{\mbox{\tiny $R\!G\!I$}}                
\newcommand{\bare}{\mbox{\tiny $B\!A\!R\!E$}}            

\def\lsim{\mathrel{\rlap{\lower4pt\hbox{\hskip1pt$\sim$}}
    \raise1pt\hbox{$<$}}}                
\def\gsim{\mathrel{\rlap{\lower4pt\hbox{\hskip1pt$\sim$}}
    \raise1pt\hbox{$>$}}}                
\PoS{PoS(LAT2005)078}

\title{
\begin{picture}(0,0)(0,0)%
   \put(0,75){\makebox(0,0)[l]{\textnormal{\normalsize DESY 05-176}}}%
   \put(0,60){\makebox(0,0)[l]{\textnormal{\normalsize Edinburgh 2005/12}}}%
   \put(0,45){\makebox(0,0)[l]{\textnormal{\normalsize Liverpool LTH 671}}}%
\end{picture}%
       A determination of the strange quark mass for unquenched clover
       fermions using the AWI}

\ShortTitle{A determination of the strange quark mass $\ldots$}

\author{Meinulf G\"ockeler$^{a}$, \speaker{Roger Horsley}$^b$,
        Alan C. Irving$^c$, Dirk Pleiter$^d$, Paul E.~L. Rakow$^c$,
        Gerrit Schierholz$^{de}$, Hinnerk St\"uben$^f$ and
        James M. Zanotti$^d$ \\
        \llap{$^a$} Institut f\"ur Theoretische Physik,
                    Universit\"at Regensburg, \\
                    D-93040 Regensburg, Germany \\
        \llap{$^b$} School of Physics, University of Edinburgh, \\
                    Edinburgh EH9 3JZ, UK \\
        \llap{$^c$} Department of Mathematical Sciences,
                    University of Liverpool, \\
                    Liverpool L69 3BX, UK \\
        \llap{$^d$} John von Neumann Institute NIC / DESY Zeuthen, \\
                    D-15738 Zeuthen, Germany \\
        \llap{$^e$} Deutsches Elektronen-Synchrotron DESY, \\
                    D-22603 Hamburg, Germany \\
        \llap{$^f$} Konrad-Zuse-Zentrum f\"ur Informationstechnik Berlin, \\
                    D-14195 Berlin, Germany \\
        E-mail: \email{meinulf.goeckeler@physik.uni-regensburg.de},
                \email{rhorsley@ph.ed.ac.uk},
                \email{aci@liverpool.ac.uk},
                \email{dirk.pleiter@desy.de},
                \email{rakow@amtp.liv.ac.uk},
                \email{gsch@mail.desy.de},
                \email{stueben@zib.de},
                \email{jzanotti@ifh.de}  }

\author{QCDSF--UKQCD Collaboration}

\abstract{Using the $O(a)$ Symanzik improved action an estimate is given for
          the strange quark mass for unquenched ($n_f=2$) QCD.
          The determination is via the axial Ward identity (AWI) and includes
          a non-perturbative evaluation of the renormalisation constant.
          Numerical results have been obtained at several lattice spacings,
          enabling the continuum limit to be taken. Results indicate a value
          for the strange quark mass (in the $\overline{MS}$-scheme
          at a scale of $2 \, \mbox{GeV}$) in the range
          $100$ -- $130 \, \mbox{MeV}$.}

\FullConference{XXIIIrd International Symposium on Lattice Field Theory \\
                 25-30 July 2005 \\
                 Trinity College, Dublin, Ireland}

\begin{document}


\section{The lattice approach}

Lattice methods allow, in principle, the complete `ab initio' calculation
of the fundamental parameters of QCD, such as quark masses. 
However quarks are not directly observable, being confined in hadrons
and are thus not asymptotic states. So to determine their mass necessitates
the use of a non-perturbative approach -- such as lattice QCD. 
In this brief article, we report on our recent results
for the strange quark mass for $2$-flavour QCD in the $\overline{MS}$-scheme
at a scale of $2 \, \mbox{GeV}$, $m_s^{\msbar}(2\, \mbox{GeV})$.
Further details can be found in \cite{gockeler05a}.


\subsection{Renormalisation group invariants}
\label{rgi}

\begin{figure}[t]
   \begin{tabular}{cc}
      \hspace*{0.15cm}
      \epsfxsize=7.00cm \epsfbox{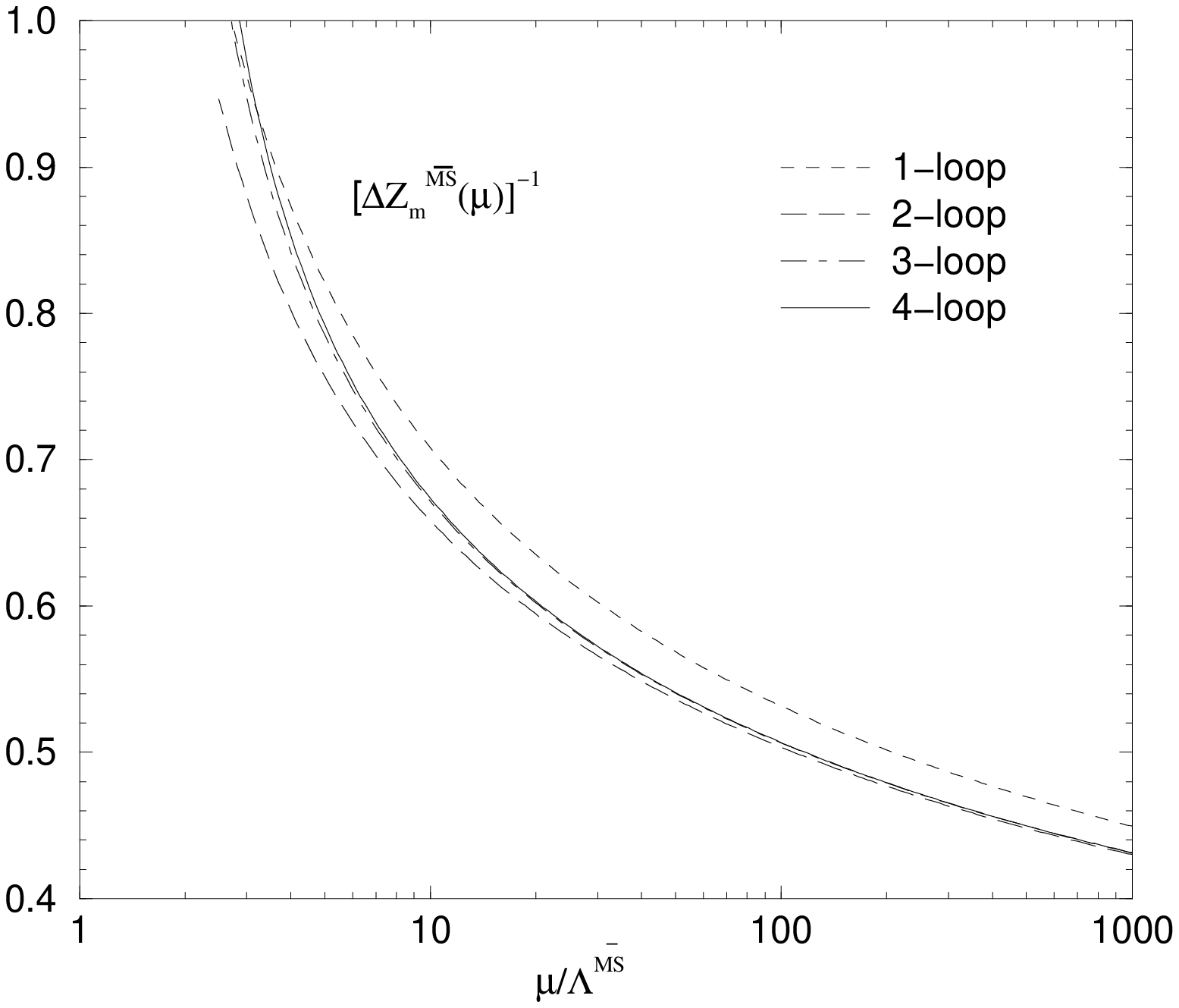}  &
      \epsfxsize=7.00cm \epsfbox{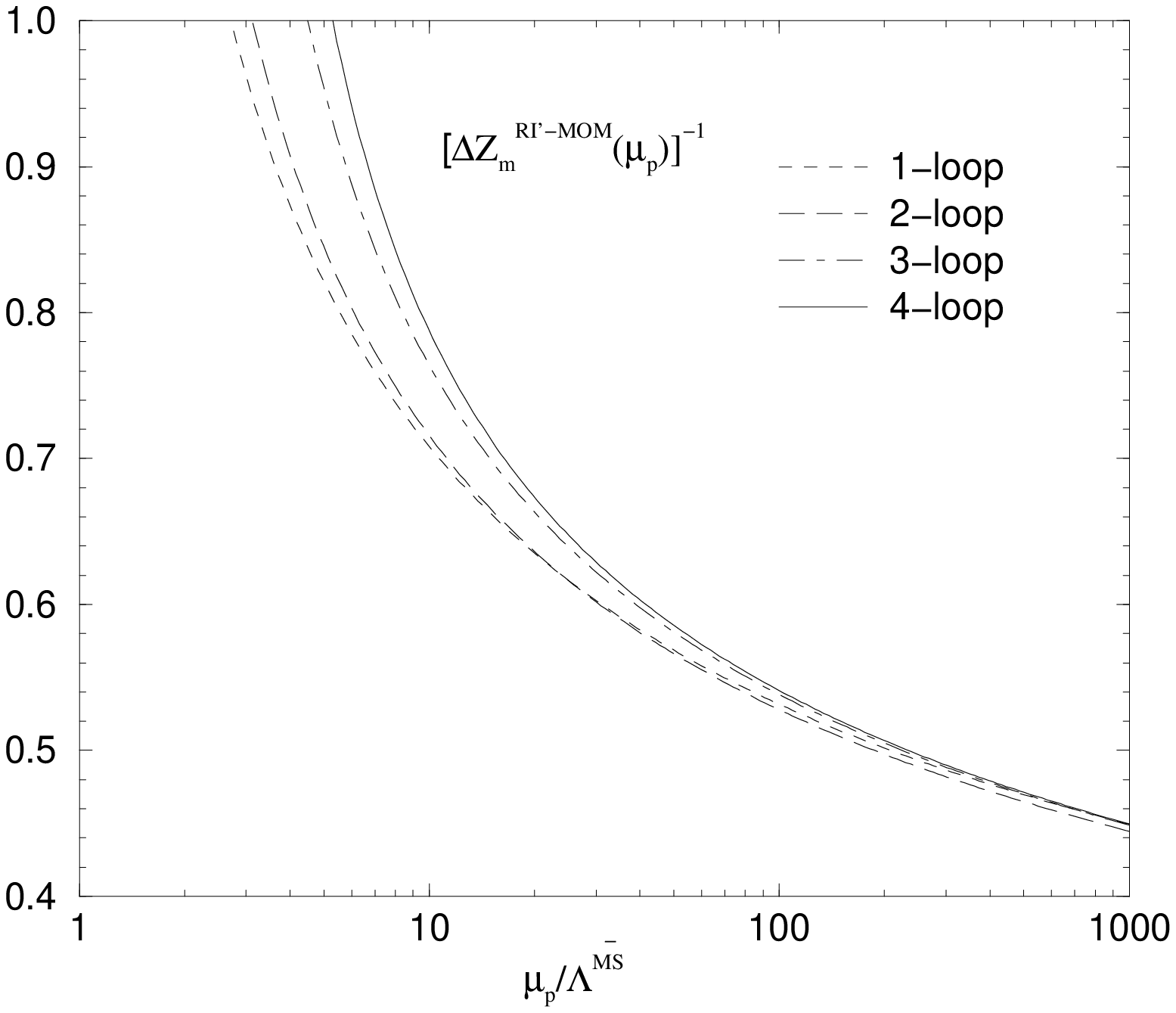}  \\
   \end{tabular}
   \caption{One-, two-, three- and four-loop results for
            $[\Delta Z_m^{\msbar}(\mu)]^{-1}$ and 
            $[\Delta Z_m^{\ripmom}(\mu_p)]^{-1}$ in units of
            $\Lambda^{\msbar}$.}
   \label{fig_Del_Zm_MSbar+RImom_nf2_muolam}
\end{figure}

Being confined, the mass of the quark, $m^{\cal S}_q(M)$, needs to be
defined by giving a scheme, ${\cal S}$ and scale $M$,
\begin{equation}
   m_q^{\cal S}(M) = Z_m^{\cal S}(M) m_q^{\bare} \,,
\end{equation}
and thus we need to find both the bare quark mass and
the renormalisation constant. An added complication is that the
$\overline{MS}$-scheme is a perturbative scheme, while more natural
schemes which allow a non-perturbative definition of the renormalisation
constants have to be used. It is thus convenient to first define a
(non-unique) renormalisation group invariant (RGI) object,
which is both scale and scheme independent by
\begin{equation}
   m_q^{\rgi} \equiv \Delta Z_m^{\cal S}(M) m^{\cal S}(M)
               \equiv Z_m^{\rgi} m_q^{\bare} \,,
\label{mrgi_msbar}
\end{equation}
where, defining the coupling constant in the chosen scheme to be always
$g^{\msbar}$ (i.e.\ expanding the $\beta^{\cal S}$ and $\gamma^{\cal S}_m$
functions in terms of $g^{\msbar}$) we have
\begin{equation}
   [\Delta Z_m^{\cal S}(M)]^{-1} = 
          \left[ 2b_0 g^{\msbar}(M)^2 \right]^{-{d_{m0}\over 2b_0}}
          \exp{\left\{ \int_0^{g^{\msbar}(M)} d\xi
          \left[ {\gamma_m^{\cal S}(\xi)
                             \over \beta^{\msbar}(\xi)} +
                 {d_{m0}\over b_0 \xi} \right] \right\} } \,.
\label{deltam_def}
\end{equation}
The $\beta^{\cal S}$ and $\gamma^{\cal S}_m$ functions (with leading
coefficients $-b_0$, $d_{m0}$ respectively) are known
perturbatively up to a certain order. In the $\overline{MS}$ scheme
the first four coefficients are known, \cite{vanritbergen97a,vermaseren97a},
and this is also true for the $\rm{RI}^\prime$-MOM scheme
\cite{martinelli94a,chetyrkin99a} (which is a suitable scheme for lattice
applications). In Fig.~\ref{fig_Del_Zm_MSbar+RImom_nf2_muolam}
we show the results of solving eq.~(\ref{deltam_def}) as a function
of the scale $M \equiv \mu$ and $M \equiv \mu_p$ for both the
$\overline{MS}$ and $\rm{RI}^\prime$-MOM schemes respectively.
We hope to use these (perturbative) results in a region where
perturbation theory has converged. $2 \, \mbox{GeV}$ corresponds to
$\mu/\Lambda^{\msbar} \sim 8$, where it would appear that the expansion
for the $\overline{MS}$-scheme has converged; for the $\rm{RI}^\prime$-MOM
scheme using a higher scale is safer (which is chosen in practice).
However, when the RGI quantity has been determined we can then easily
change from one scheme to another.
Of course these scales are in units of $\Lambda^{\msbar}$ which is
awkward to use: the standard `unit' nowadays is the force scale $r_0$.
To convert to this unit, we use the result for $r_0\Lambda^{\msbar}$
as given in \cite{gockeler05b}.


\subsection{Chiral perturbation theory}

We have generated results for $n_f = 2$ degenerate sea quarks, together
with a range of valence quark masses. Chiral perturbation theory,
$\chi$PT, has been developed for this case, \cite{bernard93a,sharpe97a}.
We have manipulated the structural form of this equation to give
an ansatz of the form
\begin{eqnarray}
   r_0m_s^{\rgi}
       &=& c^{\rgi}_a 
               \left[ (r_0m_{K^+})^2 + (r_0m_{K^0})^2 - (r_0m_{\pi^+})^2 
               \right]
                                           \nonumber \\
       & & + (c^{\rgi}_b-c^{\rgi}_d)
                \left[(r_0m_{K^+})^2 + (r_0m_{K^0})^2 \right](r_0m_{\pi^+})^2 
           + \half (c^{\rgi}_c+c^{\rgi}_d)
                \left[(r_0m_{K^+})^2 + (r_0m_{K^0})^2 \right]^2
                                           \nonumber \\
       & & - (c^{\rgi}_b+c^{\rgi}_c)(r_0m_{\pi^+})^4
           + c^{\rgi}_d (r_0m_{\pi^+})^4 \ln (r_0m_{\pi^+})^2
                                           \nonumber \\
       & & - c^{\rgi}_d \left[(r_0m_{K^+})^2 + (r_0m_{K^0})^2 \right]
                 \left[ (r_0m_{K^+})^2 + (r_0m_{K^0})^2 - (r_0m_{\pi^+})^2
                 \right] \times
                                           \nonumber \\
       & &  \hspace*{1.0in}
                 \ln \left( (r_0m_{K^+})^2 + (r_0m_{K^0})^2 - (r_0m_{\pi^+})^2
                     \right)   + \ldots \,,
\label{strange_general}
\end{eqnarray}
and
\begin{equation}
   { r_0 m_q^{\rgi} \over (r_0 m_{ps})^2 }
      = c^{\rgi}_a +
        c^{\rgi}_b (r_0 m^S_{ps})^2 +
        c^{\rgi}_c (r_0m_{ps})^2 +
        c^{\rgi}_d \left( (r_0 m^S_{ps})^2 - 2(r_0 m_{ps})^2 
                               \right) \ln (r_0 m_{ps})^2 \,.
\label{strange_degenerate}
\end{equation}
where $m_{ps}$, $m_{ps}^S$ are the valence and sea pseudoscalar masses 
respectively (both using mass degenerate quarks). The first term is
the leading order, LO, result in $\chi$PT while the
remaining terms come from the next non-leading order, NLO, in $\chi$PT.
We note that to NLO, we can determine
$c_a^{\rgi}$ and $c_i^{\rgi}$, $i = b, c, d$ using mass degenerate
quarks and then simply substitute them in eq.~(\ref{strange_general}).


\subsection{The axial Ward identity}

Approaches to determining the quark mass on the lattice are to
use the vector Ward identity, VWI (see e.g.\ \cite{gockeler04a}),
where the bare quark mass is given in terms of the hopping parameter by%
\footnote{This is valid for both valence and sea quarks. $\kappa^S_{qc}$ is
defined for fixed $\beta$ by the vanishing of the pseudoscalar
mass, i.e.\ $m_{ps}(\kappa^S_{qc},\kappa^S_{qc}) = 0$. $\kappa^S_{qc}$
has been determined in \cite{gockeler04a}.}
\begin{equation}
   m_q = {1 \over 2a}
               \left( {1\over \kappa_q} -  {1\over \kappa^S_{qc}} \right) \,,
\end{equation}
or the axial Ward identity, AWI, which is the approach employed here.
Imposing the AWI on the lattice for mass degenerate quarks, we have
\begin{equation}
   \partial_\mu {\cal A}_\mu = 2\widetilde{m}_{q} {\cal P} + O(a^2) \,,
\label{pcac_lattice}
\end{equation}
and ${\cal A}$ and ${\cal P}$ are now the $O(a)$ improved%
\footnote{The improvement term to the axial current, $\partial_\mu P$
together with improvement coefficient $c_A$, \cite{dellamorte05a}
has been included. The mass improvement terms, together
with their associated difference in improvement coefficients, $b_A$, $b_P$
appear to be small and have been ignored here.}
unrenormalised axial current and pseudoscalar density respectively
and $\widetilde{m}_{q}$ is the AWI quark mass.
So by forming two-point correlation functions
with ${\cal P}$ in the usual way, this bare quark mass can be determined.
We have found results for four $\beta$-values: 5.20, 5.25, 5.29, 5.40,
each with three sea quark masses and a variety of valence quark masses,
\cite{gockeler05a}.

\begin{figure}[t]
   \hspace{1.00in}
   \epsfxsize=8.00cm
      \epsfbox{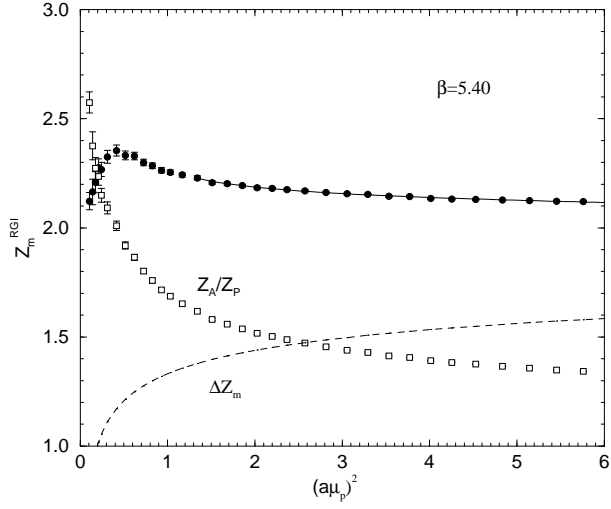}
   \caption{$\Delta Z_m^{\rgi}$ (dashed line), $Z_A/Z_P^{\rimom}$
            (empty squares) and $Z_m^{\rgi}$ (filled circles)
            for $\beta = 5.40$ (filled circles),
            together with a fit
            $F(a\mu_p) = p_1 + p_2(a\mu_p)^2 + p_3/(a\mu_p)^2$.}
   \label{fig_Zm_rgi_b5p40_ap2_050406}
\end{figure}

Furthermore upon renormalisation we have that
\begin{equation}
   Z^{\cal S}_{\tilde{m}}(M) = {Z_A \over Z^{\cal S}_P(M)} \,.
\end{equation}
As mentioned before, we use the $\rm{RI}^\prime$-MOM scheme,
to determine $Z_A$ and $Z_P^{\ripmom}$ in the
chiral limit%
\footnote{For $Z_A$ (using the sea quarks only) we make a linear
extrapolation in $am_q$, while for $Z_P^{\ripmom}$ we must subtract
out a pole in the quark mass, \cite{cudell98a}, which occurs due
to chiral symmetry breaking. We thus make a fit of the form
$(Z_P^{\ripmom})^{-1} = A_P + B_P / am_q$.}.
We now  have all the components necessary to compute $Z_m^{\rgi}$
and hence $r_0m_q^{\rgi}$. In Fig.~\ref{fig_Zm_rgi_b5p40_ap2_050406}
we show $\Delta Z^{\ripmom}_m$, $Z_A/Z_P^{\ripmom}$ and their product,
giving $Z_m^{\rgi}$ for $\beta = 5.40$. This should be independent
of the scale $(a\mu_p)^2$ at least for larger values. This seems
to be the case, we make a phenomenological fit to account for
residual $(a\mu_p)^2$ effects.

\begin{figure}[t]
   \hspace{1.00in}
   \epsfxsize=8.00cm
      \epsfbox{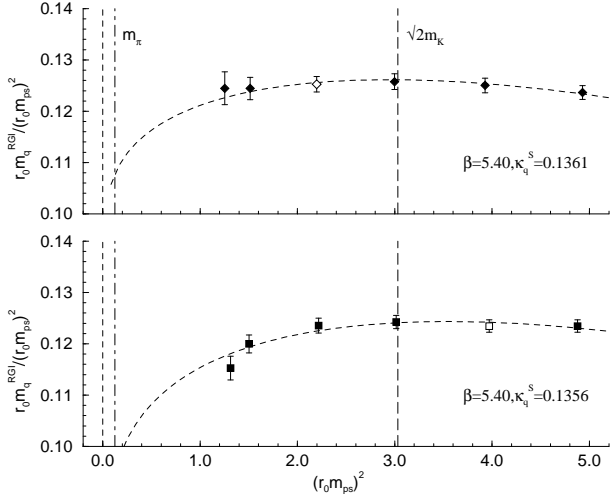}
   \caption{$r_0m_q^{\rgi}/(r_0m_{ps})^2$ against $(r_0m_{ps})^2$,
            together with a fit using eq.~(\protect\ref{strange_degenerate})
            for $\beta = 5.40$. Filled points represent valence quark results
            while unfilled points are the sea quark results.
            The dashed line (labelled `$\sqrt{2}m_K$')
            represents a fictitious particle composed of two
            strange quarks, which at LO $\chi$PT is given
            from eq.~(\protect\ref{strange_general}) by
            $\sqrt{(r_0m_{K^+})^2 + (r_0m_{K^0})^2 - (r_0m_{\pi^+})^2}$,
            while the dashed-dotted line (labelled `$m_\pi$') representing
            a pion with mass degenerate $u/d$ quark is given
            by $r_0m_{\pi^+}$.}
   \label{fig_r0mps2_r0mqRGIor0mps2_b5p40_2pic_050921}
\end{figure}

With $Z_m^{\rgi}$, we can now find $r_0m_q^{\rgi}$ and hence the
ratio $r_0m_q^{\rgi}/(r_0m_{ps})^2$, using the values
of $r_0/a$ given in \cite{gockeler05b}. In
Fig.~\ref{fig_r0mps2_r0mqRGIor0mps2_b5p40_2pic_050921} we plot this ratio
(against $(r_0m_{ps})^2$) for $\beta = 5.40$. Using
eq.~(\ref{strange_general}) to eliminate $c_a^{\rgi}$ in favour
of $r_0m_s^{\rgi}/((r_0m_{K^+})^2 + (r_0m_{K^0})^2 - (r_0m_{\pi^+})^2 )$
in eq.~(\ref{strange_degenerate}) gives $r_0m_s^{\rgi}$ directly%
\footnote{This is preferable to first determining
$c_a^{\rgi}$ and $c_i^{\rgi}$, $i=1,2,3$ by using
eq.~(\ref{strange_degenerate}) and then substituting in 
eq.~(\ref{strange_general})) as the direct fit reduces the final 
error bar on $r_0m_s^{\rgi}$.}
to NLO in our fit function.

We have restricted the quark masses to lie in the range $(r_0m_{ps})^2 < 5$,
which translates to $m_{ps} \lsim 850\, \mbox{MeV}$, which is hopefully
within the range of validity of low order $\chi$PT results. (Indeed using
$r_0/a$, rather than their chirally extrapolated values for example,
tends to give less variation in the ratio $r_0m_q^{\rgi}/(r_0m_{ps})^2$
so we expect LO $\chi$PT to be markedly dominant.)
Thus finally, for each $\beta$-value we have determined $r_0m_s^{\rgi}$
and can now perform the last extrapolation to the continuum limit.


\section{Results}

Our derivation so far, although needing a secondary quantity such as
$r_0/a$ for a unit, depends only on lattice quantities. Only at the
last stage, with our direct fit did we need to give a physical scale
to this unit. A popular choice is $r_0 = 0.5\, \mbox{fm}$.
However there are some uncertainties
in this value; our derivation using the nucleon gave
$r_0 = 0.467\, \mbox{fm}$ and so to give some idea of scale
uncertainties, we shall consider both values.
(The main change when changing the scale comes from the $r_0$s in
eq.~(\ref{strange_general}), as $m^{\rgi}_s \propto r_0$, while
changes in $\Delta Z^{\msbar}_m$ are only logarithmic.)

\begin{figure}[t]
   \hspace{1.00in}
   \epsfxsize=8.00cm
      \epsfbox{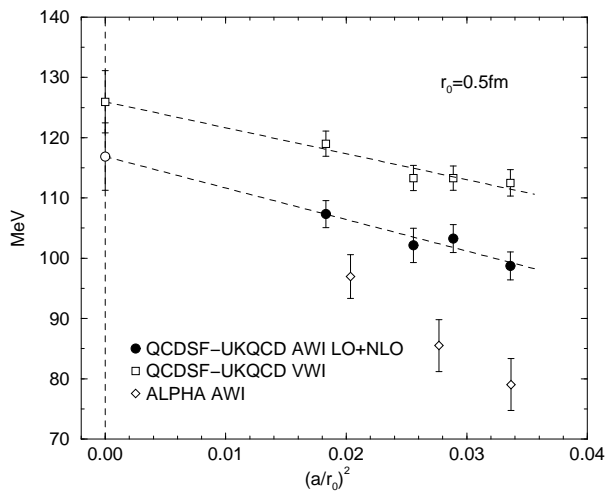}
   \caption{Results for $m_s^{\msbar}(2\,\mbox{GeV})$ (filled circles)
            versus the chirally extrapolated values of $(a/r_0)^2$ (as
            given in \protect\cite{gockeler05b}) together with a linear
            extrapolations to the continuum limit. For comparison, 
            we also give our previous result using the VWI,
            \protect\cite{gockeler04a} (open squares) and
            the ALPHA AWI determination from \cite{dellamorte05b},
            (open triangles).}
   \label{fig_msMSbar_LO+NLO_050921}
\end{figure}

Using the results from section~\ref{rgi} for 
$[\Delta Z_m^{\msbar}( 2 \, \mbox{GeV})]^{-1}$ to convert $m_s^{\rgi}$
to $m_s^{\msbar}(2\,\mbox{GeV})$ gives the results shown in
Fig.~\ref{fig_msMSbar_LO+NLO_050921}. Also shown is an 
extrapolation to continuum limit. We finally find the result
\begin{eqnarray}
   m_s^{\msbar}(2\,\mbox{GeV})
           &=& \left\{ \begin{array}{lll}
                          117(7)\,\mbox{MeV} & \mbox{for} &
                          r_0 = 0.5 \,\mbox{fm}   \\
                          111(6)\,\mbox{MeV} & \mbox{for} &
                          r_0 = 0.467 \, \mbox{fm}\\
                       \end{array}
               \right. \,,
\end{eqnarray}
where the error is statistical. This is to be compared to
our previous result using the VWI, \cite{gockeler04a},
which gave results of $126(5)\, \mbox{MeV}$, $119(5)\, \mbox{MeV}$
for $r_0 = 0.5\, \mbox{fm}$ and $0.467\, \mbox{fm}$ respectively.
We take a further systematic error on these results as being covered by
the different $r_0$ values of about $\sim 6\, \mbox{MeV}$.
Although the continuum extrapolation should be treated
with caution, it does indicate that the strange quark mass
for 2-flavour QCD lies in the region of $100$ -- $130 \, \mbox{MeV}$.


\section*{Acknowledgements}

The numerical calculations have been performed on the Hitachi SR8000 at
LRZ (Munich), on the Cray T3E at EPCC (Edinburgh)
\cite{allton01a}, on the Cray T3E at NIC (J\"ulich) and ZIB (Berlin),
as well as on the APE1000 and Quadrics at DESY (Zeuthen).
We thank all institutions.
This work has been supported in part by
the EU Integrated Infrastructure Initiative Hadron Physics (I3HP) under
contract RII3-CT-2004-506078
and by the DFG under contract FOR 465 (Forschergruppe
Gitter-Hadronen-Ph\"anomenologie).




\begin{thebibliography}{99}

\bibitem{gockeler05a}
   M. G\"ockeler \emph{et al.}, in preparation.

\bibitem{vanritbergen97a}
   T. van Ritbergen \emph{et al.},
   \emph{Phys. Lett.} {\bf B400} 379 (1997) 
   [{\tt hep-ph/9701390}].

\bibitem{vermaseren97a}
   J.~A.~M. Vermaseren \emph{et al.},
   \emph{Phys. Lett.} {\bf B405} 327 (1997) 
   [{\tt hep-ph/9703284}].

\bibitem{martinelli94a}
   G. Martinelli \emph{et al.},
   \emph{Nucl. Phys.} {\bf B445} 81 (1995)
   [{\tt hep-lat/9411010}].

\bibitem{chetyrkin99a}
   K.~G. Chetyrkin \emph{et al.},
   \emph{Nucl. Phys.} {\bf B583} 3 (2000)
   [{\tt hep-ph/9910332}].

\bibitem{gockeler05b}
   M. G\"ockeler \emph{et al.},
   {\tt hep-ph/0502212}.

\bibitem{bernard93a}
   C. Bernard \emph{et al.},
   \emph{Phys. Rev.} {\bf D49} 486 (1994)
   [{\tt hep-lat/9306005}].

\bibitem{sharpe97a}
   S.~R. Sharpe,
   \emph{Phys. Rev.} {\bf D56} 7052 (1997),
   erratum ibid. {\bf D62} (2000) 099901
   [{\tt hep-lat/9707018}].

\bibitem{gockeler04a}
    M. G\"ockeler et al.,
    {\tt hep-ph/0409312}.

\bibitem{dellamorte05a}
   M. Della Morte \emph{et al.},
   \emph{JHEP} {\bf 0503} 029 (2005)
   [{\tt hep-lat/0503003}].


\bibitem{cudell98a}
   J.~R. Cudell \emph{et al.},
   \emph{Phys. Lett.} {\bf B454} 105 (1999)
   [{\tt hep-lat/9810058}].

\bibitem{dellamorte05b}
   M. Della Morte \emph{et al.},
   {\tt hep-lat/0507035}.

\bibitem{allton01a}
   C.~R. Allton \emph{et al.},
   \emph{Phys. Rev.} {\bf D65 } 054502 (2002) [{\tt hep-lat/0107021}].

\end{thebibliography}
\end{document}